\documentclass[]{aastex631}

\usepackage{natbib}
\usepackage{color}
\usepackage{soul}
\usepackage{csquotes}

\usepackage{graphicx} 
\usepackage{array} 

\newcommand{\bali}{\begin{align}}

\newcommand{\eali}{\end{align}}

\def\br{{\bf r}}
\def\br'{{\bf r'}}

\def\N{\mathbb{N}}

\def\Z{\mathbb{Z}}
\def\Q{\mathbb{Q}}

\def\D{\bf D}

\def\det{\mbox{det}\;}

\def\r{{\bf r}}

\def\R{{\bf R}}

\def\x{x}
\def\y{y}
\def\z{z}

\def\zu{{\bf{\hat z}}}

\def\k{{\bf k}}
\def\h{{\bf h}}

\def\d{{\rm d}}

\def\n{{\bf{\hat n}}}

\def\C{{\cal C}}
\def\L{{\cal L}}
\def\S{{\cal S}}
\def\V{{\cal V}}
\def\B{{\bf B}}

\def\b{{\bf b}}

\def\E{{\bf E}}

\def\V{{\bf V}}

\def\j{{\bf j}}

\def\curl{\mathop{\bf{\nabla}\times}}
\def\R{{\bf R}}
\def\N{{\bf N}}
\def\d{{\rm d}}
\def\B{{\bf B}}
\def\b{{\bf b}}
\def\A{{\bf A}}
\def\J{{\bf J}}
\def\X{{\bf X}}
\def\q{{\bf q}}
\def\r{{\bf r}}

\def\B{ {\bf B} }

\def\b{{\bf b}}

\def\R{ {\bf R}  }

\shorttitle{Twisted Flux Ropes in the Quiet Sun}
\shortauthors{Amari et al.}
\graphicspath{{./}{figures/}}
\begin{document}

\title{The Ubiquity of Twisted Flux Ropes in the Quiet Sun
\footnote{Released on March, 1st, 2021}}

\author[0000-0002-2115-1284]{Tahar Amari}
\affiliation{CNRS, Centre de Physique Th\'eorique de l'Ecole
Polytechnique, IPP, \\
F-91128 Palaiseau Cedex, FRANCE}

\author{Aur\'elien Canou }
\affiliation{CNRS, Centre de Physique Th\'eorique de l'Ecole
Polytechnique, IPP, \\
F-91128 Palaiseau Cedex, FRANCE}

%\collaboration{6}{(AAS Journals Data Editors)}
\author{Marco Velli }
\affiliation{Department of Earth, Planetary Space Sciences
University of California, Los Angeles, CA 90095-1567.USA.
}

\author{Zoran Mikic}
\affil{Predictive Science Inc. 9990 Mesa Rim Rd, Suite 170
San Diego, CA 92121. USA.
}

\author{Frederic Alauzet }
\affiliation{INRIA Saclay Ile-de-France, Projet Gamma3, \\
1, rue Honoré d'Estienne d'Orves, 91126 Palaiseau,  France }

\author{Eric Buchlin }
\affiliation{Universit\'e  Paris-Saclay, CNRS, Institut d'Astrophysique Spatiale
Orsay France
}
\author{Jean-Fran\c cois  Luciani}
\affiliation{CNRS, Centre de Physique Th\'eorique de l'Ecole
Polytechnique, IPP, \\
F-91128 Palaiseau Cedex, FRANCE}

\author{Jean -Jacques Aly }
\affiliation{AIM - Unit\'e Mixte de Recherche CEA - CNRS \\
 Universit\'e Paris VII - UMR no 7158,  \\
Centre d'Etudes de Saclay, F-91191 Gif sur Yvette Cedex,  FRANCE}

\author{Lucas A. Tarr }
\affiliation{National Solar Observatory.22 Ohia Ku St
Makawao, HI 96768.USA.}

%% Note that the \and command from previous versions of AASTeX is now
%% depreciated in this version as it is no longer necessary. AASTeX 
%% automatically takes care of all commas and "and"s between authors names.

%% AASTeX 6.31 has the new \collaboration and \nocollaboration commands to
%% provide the collaboration status of a group of authors. These commands 
%% can be used either before or after the list of corresponding authors. The
%% argument for \collaboration is the collaboration identifier. Authors are
%% encouraged to surround collaboration identifiers with ()s. The 
%% \nocollaboration command takes no argument and exists to indicate that
%% the nearby authors are not part of surrounding collaborations.

%% Mark off the abstract in the ``abstract'' environment. 
\begin{abstract}

Models and observations have demonstrated that Twisted Flux Ropes (TFRs) play a significant role in the structure and eruptive dynamics of active regions. Their role in the dynamics of the quiet Sun atmosphere on has remained elusive, their fundamental relevance emerging mainly from theoretical models  (Amari et al. 2015), showing that they form and erupt as a result of flux cancellation. Here HINODE high-resolution photospheric vector magnetic field measurements are integrated with advanced environment reconstruction models: TFRs develop on various scales and are associated with the appearance of mesospots. The developing TFRs contain sufficient free magnetic energy to match the requirements of the recently observed "campfires" discovered by Solar Orbiter in the quiet Sun. The free magnetic energy is found to be large enough to trigger eruptions while the magnetic twist large enough to trigger confined eruptions, heating the atmosphere. TFRs are also connected to larger scale magnetic fields such as supergranulation loops, allowing the generation of Alfv\'en  waves at the top of the chromosphere that can propagate along them. High-resolution magnetohydrodynamic simulations, incorporating subsurface dynamo activity at an unprecedented 30 km spatial resolution, confirm that TFRs are ubiquitous products of the permanent small scale dynamo engine that feeds their formation, destabilization, eruption via flux emergence, submergence and cancellation of their chromospheric feet, similar to the dynamics driving large scale eruptive events. Future investigations, especially with the Daniel K. Inouye Solar Telescope (DKIST) and Solar Orbiter will deepen our understanding of TFRs in the context of atmospheric heating.\\
\end{abstract}

%% Keywords should appear after the \end{abstract} command. 
%% The AAS Journals now uses Unified Astronomy Thesaurus concepts:
%% https://astrothesaurus.org
%% You will be asked to selected these concepts during the submission process
%% but this old "keyword" functionality is maintained in case authors want
%% to include these concepts in their preprints.

%\keywords{Classical Novae (251) --- Ultraviolet astronomy(1736) --- History of astronomy(1868) --- Interdisciplinary astronomy(804)}

%% From the front matter, we move on to the body of the paper.
%% Sections are demarcated by \section and \subsection, respectively.
%% Observe the use of the LaTeX \label
%% command after the \subsection to give a symbolic KEY to the
%% subsection for cross-referencing in a \ref command.
%% You can use LaTeX's \ref and \label commands to keep track of
%% cross-references to sections, equations, tables, and figures.
%% That way, if you change the order of any elements, LaTeX will
%% automatically renumber them.
%%
%% We recommend that authors also use the natbib \citep
%% and \citet commands to identify citations.  The citations are
%% tied to the reference list via symbolic KEYs. The KEY corresponds
%% to the KEY in the \bibitem in the reference list below. 

%\begin{document}

\section{Introduction}
\label{sec:LIntro}

The quiet Sun, defined as a region of the outer solar atmosphere not dominated by the strong magnetic activity of sunspots and active regions,  constitutes the largest portion of the solar surface, persisting throughout the solar cycle. However, since the observations from SOHO it has been shown that the quiet Sun is also incredibly dynamic, exhibiting a wide range of bursty structures, from the spicules well known from H-alpha observations to mottles, blinkers, jets,  tornadoes, and the recently observed Solar Orbiter EUV brightenings nicknamed "campfires" 
% EB On ne prétend pas que les campfires sont un type différent de structures que celles observées à plus grande échelle
\citep{BerghamsAuetal2021}, extending to various heights from the chromosphere to the core corona. Understanding how these phenomena are related to underlaying structures and their contribution  to the heating of the solar atmosphere -- and reconciling the heating with observations at various altitudes -- remain major challenges \citep{Aschwandenetal2007}.

Although it is not known whether the physical mechanisms responsible for the thermal structuring of the Quiet Sun's atmosphere begin immediately above the temperature minimum, they are certainly active at an altitude of 4000 km. While this zone remotely observed by  the Solar Orbiter,  is found very active, it remains out of reach of  in-situ measurements for the foreseeable future.  Characterizing the underlying  magnetic structures associated with heating mechanisms, such as their length scales, is thus of paramount importance.

A fundamental discovery on the nature of the quiet Sun was the observation that small-scale magnetic fields \citep{Lites2008, TrujilloBueno2004}, and particularly  horizontal ones, dominate its structure. It is thus important to  investigate how such small scale  internetwork magnetic fields are related to the larger supergranular network magnetic fields emerging above \citep{Gosicetal2024}.

Over the past decades, Twisted Flux Ropes (TFRs)  have been recognized as crucial structures for understanding the magnetic structure and support of prominences \citep{KuperusRaadu1974,AnzerPri85,AmariAl89, DemoulinPriest88,VanBallegooijenMa89,AulanierDe98} and the mechanisms behind solar eruptions \citep{AlyAmari85,Low01,PriestForbes02,AmariAl09,TorokKl05, AmariCaAl2014,Aulanier2014,AmariAlCa2018, KliemTo06,Toroketal2018}. The formation and evolution of a TFR towards eruption can occur and be modeled in different ways, including the emergence of a pre-existing TFR  from the convection zone \citep{Fan2022}. Numerous remarkable studies driven by numerical simulations have been performed, considering different physical ingredients, either idealistically by coupling different zones of the atmosphere \citep{AmariLuAl04,AmariLuAl05,FanGibson04,FanGib07} or with more realistic models including the different layers \citep{ArchontisTor2008, Cheungetal2019, LeakeLinAn2022}. Despite the difficulties in observing convection zone magnetic fields, there is indirect evidence supporting the emergence of such formed TFR \citep{LopenzFuentesDeManVan00,MacTaggartetal2021}.

On the other hand, reconstruction of the coronal magnetic field from vector magnetograms above active regions shows the existence and formation of sheared magnetic arcades and TFRs prior to eruption, with the former  being a prerequisite  stage during  the slow evolution leading to TFR and eruption \citep{Patsourakos2020}. Whether the TFR is fully formed before or during the eruption remains a matter of debate.   All these studies (theoretical and observational) of active regions have shown that the photospheric transverse magnetic field is a key ingredient for the energization of highly sheared magnetic field structures and TFRs.

The importance of TFRs for the structure and dynamics of the quiet Sun has only been appreciated more recently, as initially suggested by the somewhat idealized numerical simulations of \citep{AmariLuAl2015}. It was shown that reconnection provides chromospheric heating as recently found also in active regions \citep{Bose2024}, while in upper layers waves also play a role. The simulations demonstrate that TFRs evolve in response to magnetic flux changes occurring at their chromospheric feet, which eventually leads to their eruption into upper layers, as in some models of coronal mass ejections. Recently  more realistic simulations  \citep{ChenPeter2021, MartinezSikora2019, Robinson2023} have also shown that TFRs  may form. 

Given the importance of TFRs, it is crucial to examine data in the quiet Sun. While SDO/HMI vector magnetograms have been intensively used in the case of active regions, in Section~\ref{sec:reconstruction} the present paper makes use  of the  vector magnetic field data from Hinode, which enables higher spatial resolution and lower noise in particular  for the transverse component \citep{Lites2008,Regnier2008}.
The focus is on small-scale vector magnetic field measurements of the quiet Sun by Hinode  \citep{Lites2008}. The usual assumption of  current-free (potential) magnetic field configuration  \citep{Schrijver2003} is relaxed in Section~\ref{sec:simulation} to allow for force-free magnetic fields using the XTRAPOL model 
% EB besoin de dire aussi que c'est du _nonlinear_ force free?
\citep{AmariAl10} to compute the magnetic structures  emerging from the photospheric field measurement. These yield ubiquitous small-scale TFRs in the quiet Sun region. In addition  results from a new high resolution (at 30 km) MHD  numerical simulation of a small  portion of the quiet Sun are also shown that  support the reconstruction findings:  TFRs,  produced by the sub-photospheric dynamo with emergence and cancellation, clearly populate the whole quiet Sun, reaching into the chromosphere and  above \citep{Gosicetal2024}.

\section{Observations and Reconstruction of the magnetic environment of a typical quiet Sun region}
\label{sec:reconstruction}

\begin{figure}
%\plotone{flines_t800_lp_800_5_1}
\epsscale{1.}
%\plotone{fig1.eps}
\plotone{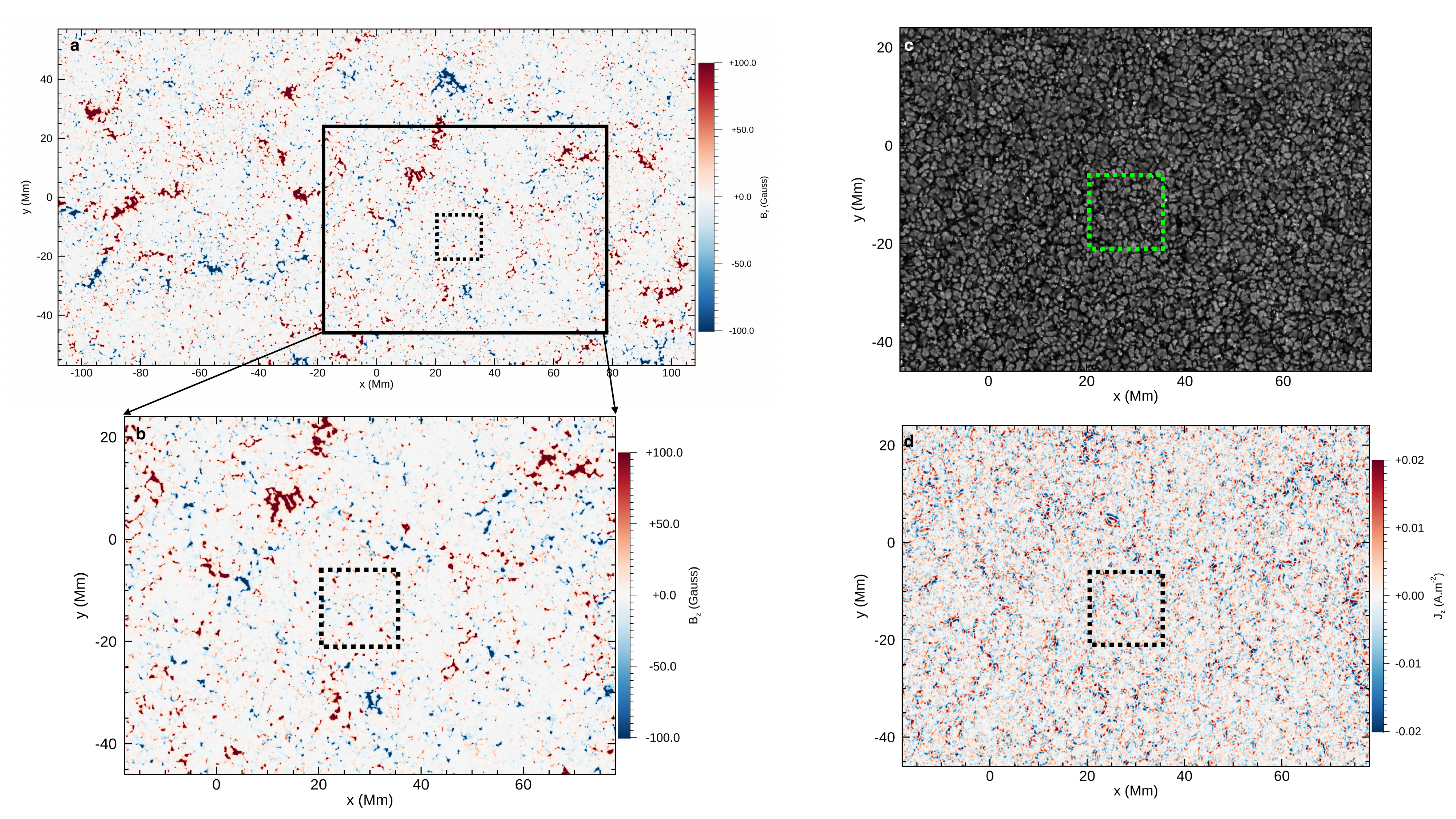}
\caption{HINODE/SOT high resolution photospheric magnetic map used to build up adapted resolution boundary conditions. a. The normal component $B_z$ of the magnetic field for the full-size magnetogram ($216 \times 114$\,Mm$^2$),from which we extract a subdomain (solid line rectangle)  of $96 \times 70$\,Mm$^2$ corresponding to the domain above which coronal magnetic field is reconstructed of, and within which a smaller one (dashed line) corresponds to the domain of simulation ($15 \times 15$\,Mm$^2$). b. Zoom on the normal component $B_z$ for the reconstruction domain. c. Continuum image for the reconstruction domain exhibiting the granulation scale. d. The normal component $J_z$ of the electric current for the reconstruction domain.
\label{fig_reconst_bzjz} }
\end{figure}

To address this problem a region of the quiet Sun located near the disk center on 2007/03/10 \citep{Lites2008} was targeted. The Spectro-Polarimeter (SP) of the Solar Optical Telescope \citep[SOT,][]{Tsuneta2008} on board the satellite Hinode \citep{Kosugi2007} measured the Stokes parameters in normal  mode providing a high resolution map with pixel size of (0.148570 ; 0.159992) arcsecs in the (x,y) helioprojective Cartesian coordinates. The level-2 Hinode/SOT/SP  data%
\footnote{The L2-data are available at \url{https://csac.hao.ucar.edu/sp_data.php}. They have already been inverted from Stokes parameters using the Merlin algorithm \citep{Skumanichetal1987}.
}
 contain the continuum field, the norm of $\B$, the inclination and azimuth angles; the azimuthal angle being known to suffer from the {180$^{\circ}$}-amibiguity. The azimuthal ambiguity is resolved with the Minimum Energy Method \citep{Lekaetal2009} applied a thousand times with different seeds to  determine statistically the most probable solution. These data are then converted to Cartesian magnetic field components and remapped to a cartesian grid \citep{Garyetal1990} with pixel sizes $(dx ; dy)$ of $(0.1 ; 0.1)$\,Mm covering an area of  $216 \times 114$\,Mm$^2$. The current density  $J_z\,\zu = \curl \B_{T}$  and $\alpha = J_z/B_z$ are computed on that grid as long as $\|B_{T}\|>10$\,G and $|B_z| > 5$\,G, respectively.  An additional 5-point gaussian smoothing is applied to get rid of abrupt variations.   Figure~\ref{fig_reconst_bzjz} presents the result of this preparation.

It is worth noticing that the  region of interest is associated with a period of the declining phase of solar cycle 23, during which no active region was present.

\begin{figure}
\plotone{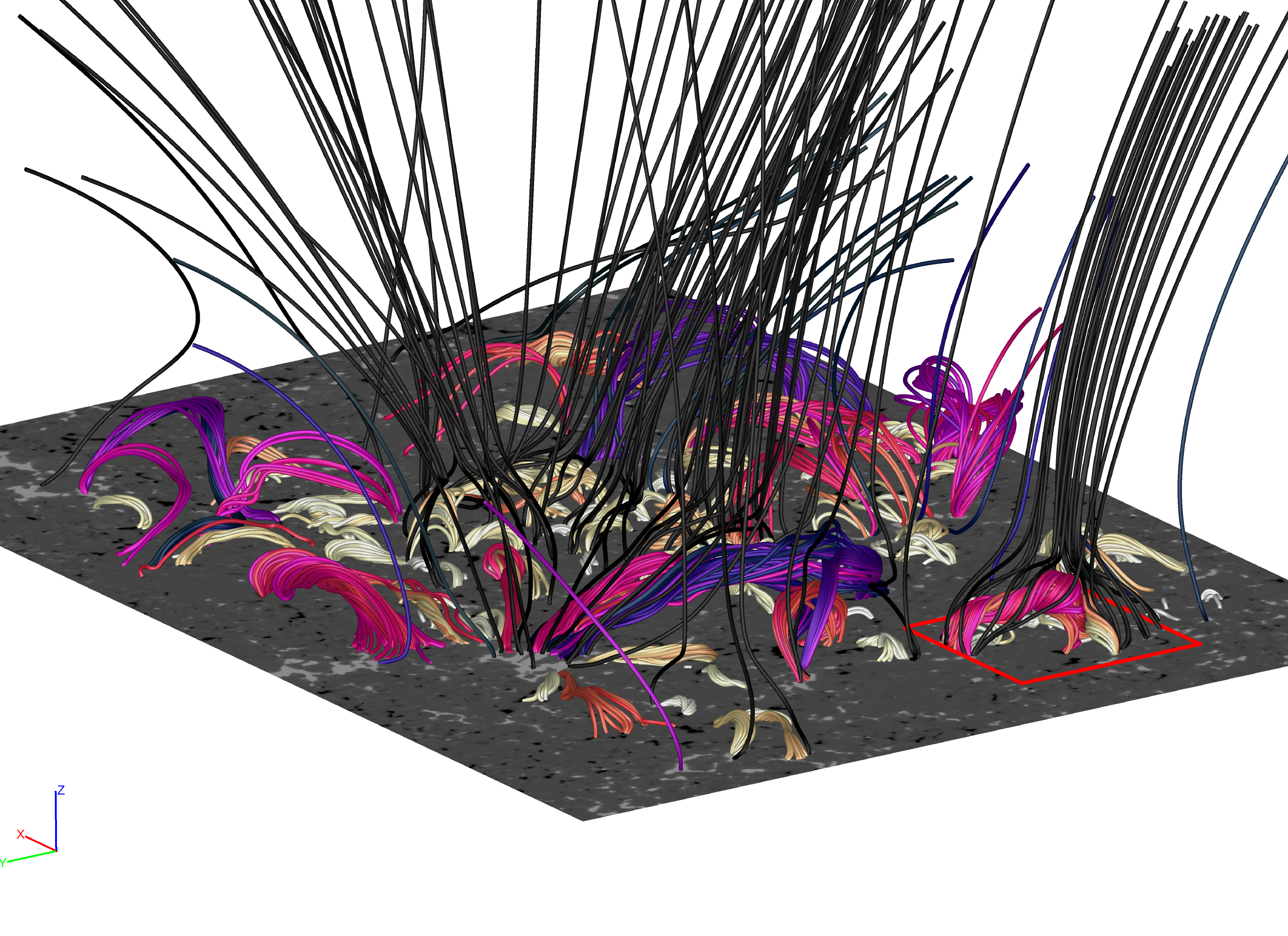}
\epsscale{1.}
%\plotone{fig2.eps}
\caption{Selected field lines of the magnetic environment above the photosphere using the force-free model XTRAPOL. Twisted Flux Ropes stand around mesospots, both below and above the transition regions. The red rectangle shown on the right, features the same box size  as the one of the simulation box in Figure~\ref{fig_simulation_tfr}.} 
\label{fig_reconst_tfrs}
\end{figure}

\begin{figure}
\plotone{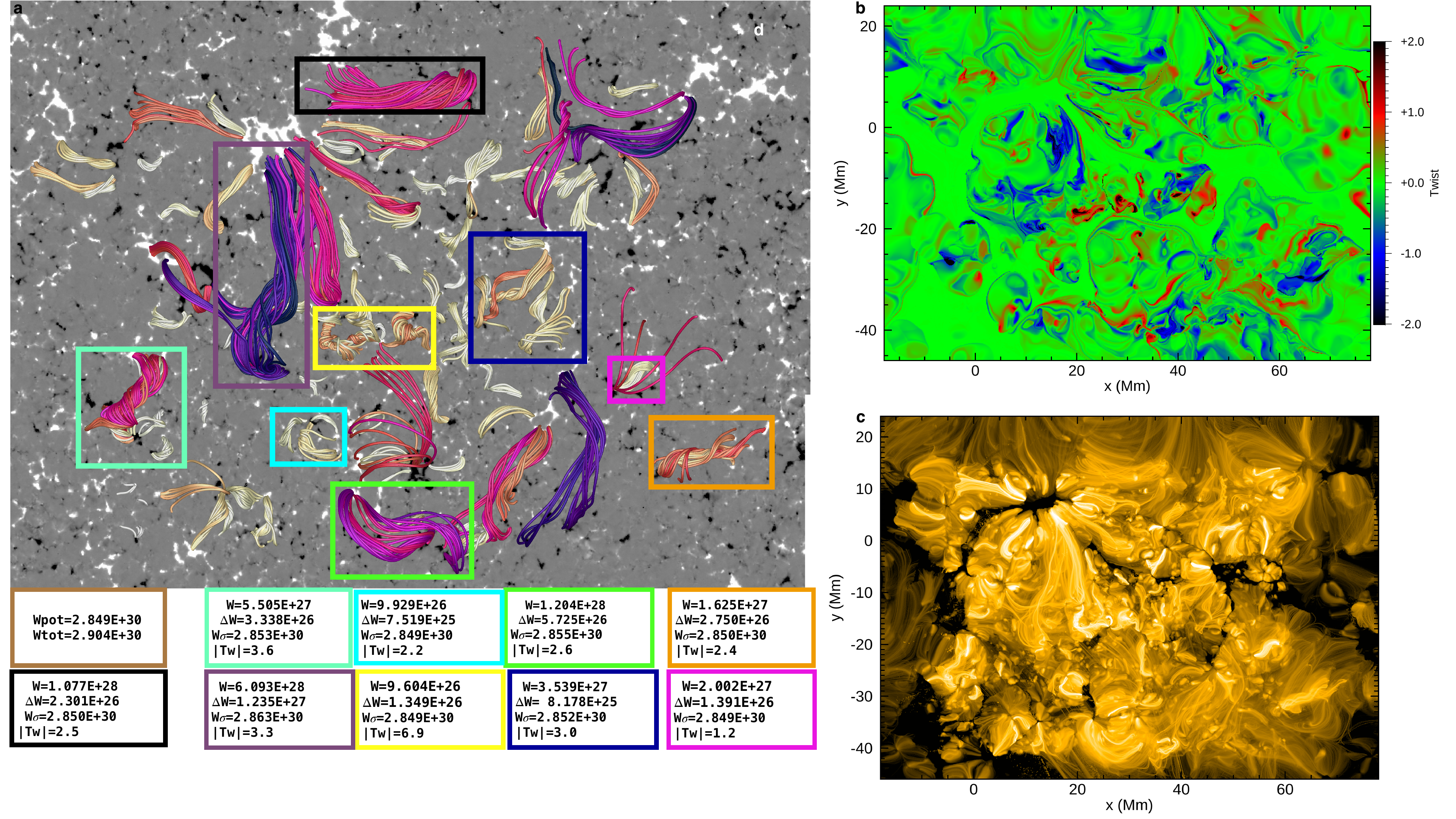}
\epsscale{1.}
%\plotone{fig3.eps}
\caption{a. Boxes enclosing TFRs used to compute some of their characteristics. Each box can contain one or a few TFRs when difficult to isolate. b. Magnetic twist at $z = 1, \text{Mm}$. c. Synthetic emissivity from the electric current density. Below  panel a are listed values of: $W_{\text{tot}}$ and $W_{\text{pot}}$, the magnetic energy of the reconstructed solution and the current-free one for the entire magnetogram; $W$, the magnetic energy contained in each box; $\Delta W$, the free magnetic energy (compared to the current-free field); $W_{\sigma}$, the semi-open field energy associated with the opening of each box; $|Tw|$, the magnetic twist of the TFR in each box (or the maximum when several are present). All magnetic energies are expressed in erg.
\label{fig_reconst_energy_boxes_twist_emissivity}}
\end{figure}

The first few thousand kilometers above the photosphere are extremely dynamic, but as the plasma $\beta$ decreases quickly with height fundamental insights can be obtained by assuming the atmospheric domain to be filled with a low-$\beta$ slightly resistive and viscous plasma.  Then {$\bf{B}$} obeys the standard force-free equations \citep{AmariCaAl2014} and can be reconstructed from the observed photospheric field with appropriate boundary conditions at the top and the sides of the domain.   The force-free field is obtained using the numerical model XTRAPOL \citep{AmariAl10} that integrates the equilibrium equations up to an altitude of 80\,Mm on a non uniformly distributed grid with a total of 321 points in the vertical direction. Thereafter and without loss of generality we focus on a sub volume of {$96 \times 70 \times 80$\,Mm$^3$ that embraces granulation and  supergranulation. The horizontal spatial resolution is uniform with $dx=dy=100$\,km, while vertically constant $dz=50$\,km up to  8\,Mm,  and then increasing to reach 750\,km at 80\,Mm.

The magnetic  configuration differs substantially from that of the potential field that has the smallest magnetic energy compatible with the flux distribution at the lower photospheric boundary \citep{Schrijver2003}.
One observes field lines that meander and tangle into a number of TFRs, shown in Figure~\ref{fig_reconst_tfrs}. The characteristic TFR length is associated with an intermediate coherent scale we have previous dubbed mesospots  \citep [see section below for the simulation]{Bushby2014,AmariLuAl2015}.
%This is better seen by showing cut of the normal component of the magnetic field at typical height of  middle-upper chromosphere (at 1.5\,Mm above the surface) which reveal those mesospots . 
% Due to the presence of electric currents, 
% je trouve que "electriic current" est un peu ambigue ? non ?
The presence of a current density $\bf{j}$ in the whole volume means that the magnetic field energy in the extrapolted field is much greater than the energy of the associated current-free (potential) magnetic field having the same normal componentat the photosphere).

The TFRs recovered from XTRAPOL have typical apex heights around or  below  7\,Mm, similar to the heights of the so-called "campfires" \citep {Zukhovetal2021} recently observed by Solar Orbiter. To better analyze the magnetic field configuration and its potential for energetic dynamics, a simplified segmentation of the domain was performed, using boxes to encompass single TFRs when possible or several  TFRs where the field configuration included close or interlaced twisted fields. In this way more localized estimates of the distribution of the magnetic energy inside the reconstructed domain, and specifically associated with interesting TFRs (Figure~\ref{fig_reconst_energy_boxes_twist_emissivity}) were obtained.  The values recovered match those estimated by \cite{Panesaretal2021} to sustain the losses from "campfires" ($10^{26}-10^{27}$ erg), which require an average magnetic field of the order of 20 G. As shown in  Figure~\ref{fig_reconst_energy_boxes_twist_emissivity}, the values found in the boxes are of same order and even larger since  $9.6\;10^{26}\,\mbox{erg}  \le  W \le 6.1\;10^{28}\,\mbox{erg}$,  while the difference between the magnetic energy and the current-free one is in the interval $7.5\;10^{25}\,\mbox{erg} \le  \Delta W \le 1.2\;10^{27}\,\mbox{ erg}$. Furthermore we find  that the magnetic energy contained in each box, is larger than that required to trigger non confined eruptions, when compared to the energy of the  semi-open field  \cite[see][]{AmariCaAl2014}. The magnetic  twist in individual TFRs is also found to be  large, with many values above $\|Tw\|=2$, favourable to trigger confined eruptions via the kink instability \citep[][and references  therein; see bottom panel of Figure~\ref{fig_reconst_energy_boxes_twist_emissivity}]{AmariAlCa2018}.

The magnetic topology is  highly complex,   splitting the domain into many magnetic cells delimited by singular surfaces, the so-called separatrices.  This  is achieved by  computing the so called squashing  factor $Q$  \citep{Titov2007}, exhibiting various  delimited regions, both in the  sense of exact separatrices surfaces \citep{Parker1983}, due  to the highly multipolar distribution of magnetic field,  and also of  Quasi Separatrix Layers \citep[QSLs;][]{Demoulin1996}, where $Q$ is very large. These QSLs  contain separated bundles of field lines, becoming  zones of  observed emission, during the evolution of the system as a result of magnetic reconnection.
% here shown using squashing factor (Figure~\ref{fig_reconst_qls}). \\
A synthetic emissivity  is  computed for the extrapolated volume using a method introduced  for models of this type  which do not evolve the complete plasma dynamics, including thermodynamic quantities such as density and temperatures that are necessary for
atomic physics emissivity calculations \citep{Cheungetal2012,AmariAlCa2018} such as can be obtained using the CHIANTI package \citep{CHIANTI1,CHIANTI10}. 
% EB pour suivre ce que CHIANTI demande de citer: le 1er article et celui qui est utilisé (ou le dernier)
The method is based on a preliminary calculation of millions of field lines, along which the electric current density is then integrated. An emissivity proxy is then defined for each cell of the atmosphere by summing the contributions of each line that passes through that cell.
The emissivity of the  reconstructed environment is shown in  Figure~ \ref{fig_reconst_energy_boxes_twist_emissivity}-c. Hot spots associated with the large electric currents flowing along lines are clearly visible and  bear a striking resemblance to "campfires".

The examined TFRs share several typical properties of their active region counterparts, and may therefore be associated with eruptions in the neighbourhood of mesospots, triggered by mechanisms (such as flux cancellation) similar to those at work in large scale eruptive events \citep[see][and the section below about numerical simulations]{AmariLuAl2015}.

Such TFRs are also coupled to the large scale magnetic fields including super-granulation loops.  As shown in Figure~\ref{fig_reconst_tfrs},  the TFR on the bottom right side is located within a dome-shaped structure that opens to large scales, resembling the classical configuration involved in jets. An eruption of this  TFR (caused by its large twist/energy) would affect the corona directly through spicule type activity and waves and jets, as already seen  in  the numerical calculation of \cite {AmariLuAl2015} but also characteristic of many  observations  of eruptive active region magnetic configurations   \citep{Schmieder2022}. Owing to those connections with larger scale magnetic field, the perturbations resulting from their possible eruption and reconnection, happening near the top of the chromosphere, generate Alfv\'en waves that can propagate along the vertical tubes and contribute to heating higher up, which is not considered here.  It is worth noting that unlike the typical configurations in \cite {AmariLuAl2015}, the large scale supergranulation scale loops  carry only a small amount of twist  in the present force free  reconstruction. This is coherent with the fact that equilibrium solutions cannot catch waves. As can be seen in the red rectangle representing the typical size of the MHD simulation $15 \times 15$\,Mm$^{2}$) in Figure~\ref{fig_simulation_tfr}, the scale of the TFRs which occupies a significant part of the box and which we had described as mesoscale, can be either smaller or larger than this box, but always above the granular scale.

\section{ Numerical Simulation of the emergent magnetic field structure of a quiet Sun region}
\label{sec:simulation}

Here we illustrate the formation of dynamic TFRs in the Quiet Sun via numerical simulation. To do so, we build upon the established result that the key source of energy is the small-scale magnetic field, as mentioned earlier in the introduction \citep{Lites2008, TrujilloBueno2004}.  Our approach is motivated by the aim {to explore of understanding the necessary conditions that result from maintaining a hot atmosphere with a sharp transition region, statistically steady constant in time, and of determining the resulting structures that are formed and compatible with the atmosphere}. Full details of the method, initially introduced in \cite{AmariLuAl08} and further improved, are provided in \cite{AmariLuAl2015}, including a comparison of the simulated dynamo-generated surface magnetic field with the fields distribution inferred from observations of scattering polarization in atomic and molecular lines. We briefly recall here the main points of the approach and improvements made.

The magnetic field is assumed to be created by a subsurface small-scale dynamo operating in the upper 1.5\,Mm of the convection zone, where the plasma is taken to obey the incompressible Boussinesq MHD equations. This layer is coupled with an atmospheric region described by a different MHD model that includes a photosphere and a chromosphere with respective thicknesses of 0.5\,Mm and 1.5\,Mm, as in the Sun though the observed extent of the chromosphere fluctuates and may be up to 5\,Mm high e.g. in coronal holes}, and a corona extending up to 30\,Mm. We start from an equilibrium atmosphere solution of the hydrostatic equation with a density profile differing slightly from VAL data. 
% EB missing reference for the VAL model
Coupling between the underlying dynamo and the atmosphere is achieved through the Resistive Layer Model {\citep{AmariLuAl2015}. The atmosphere is initially taken to be in equilibrium, and its temperature is kept constant in time, in line with our aim of maintaining the sharp variations and the saturated dynamo.

%We thus do  not consider  the thermodynamic and radiative response of the plasma unlike in  realistic  approach adopted as by others such as Bifrost33, MURaM34, CO5BOLD35, PENCIL36, or Stagger37. This way we also think that it is easier to keep a sustained, saturated  dynamo while   maintaining the whole atmospheric temperature including in particular  the  sharp region.

A few differences with respect to \cite{AmariLuAl2015} are worth noting.  The code has been improved with MPI parallelisation, allowing a larger spatial resolution of $512 \times 512 \times 400$ for the same horizontal domain, but now extending vertically to 30\,Mm. The fully implicit numerical scheme provides better resolution for the resistive part, and a higher-order scheme for our numerical MHD code METEOSOL  has been implemented since its last version \citep{AmariLuAl2015}. The mesh is horizontally uniform and vertically uniform in the convection zone, while non-uniform in the atmosphere, with a typical spatial resolution of 0.03\,Mm.

After reaching a saturated dynamo, as in the lower resolution simulation, we observed the emergence and submergence of structures. The temperature is structured at the granulation scale while the magnetic field exhibits both granulation scale and mesoscale features, clearly showing that flux cancellation of magnetic concentrations of opposite polarities occurs frequently throughout the evolution. Although on the photosphere the magnetic field appears to be organized at the granulation scale of fluid motions, it also exhibits more persistent (30 minutes lifetime) mesoscale magnetic flux concentrations that play an important role in the system evolution, which we previously called "mesospots" by analogy with active region sunspots with sizes several times that of a  granulation cell.

% The  time averaged Poynting flux, whose divergence controls the transfer of energy from the magnetic field to the plasma,  can be used as a heating input in mean atmospheric models, is found to be consistent with the flux required at the base of the chromosphere (4500 W/m2) and at the transition region (300 W/m2).
%
Higher up, as already discovered in \cite{AmariLuAl2015}, coherent structures that are no longer confined generate emerging motions by expanding into the region above. These structures are present at 1.5\,Mm above the surface, where they appear as coherent magnetic flux tubes with a width of about 1--2\,Mm and a mean intensity of about 20 G. Among these structures, TFRs appear (Figure~\ref{fig_simulation_tfr}), with a physics dominated by the relatively stable mesospots. At the periphery of these mesospots, dynamic current sheets persistently structure the plasma and are associated with magnetic complexity.
%(Figure~\ref{fig_simulation_qsl}) .
TFRs associated with these mesospots form, and eruptions in the neighbourhood of the mesospots are triggered by mechanisms such as magnetic flux cancellation, similar to those at work in large-scale eruptive events, but occurring at higher altitudes and involving their chromospheric feet. As seen in Figure~\ref{fig_simulation_tfr}, the horizontal size of the domain corresponds to the smallest rectangle drawn in green in Figure~\ref{fig_reconst_bzjz}, exhibiting several TFRs, which supports the widespread presence of these structures across the entire quiet Sun.
%(Figure~\ref{fig_simulation_tfr_eruption}).  
%Synthetic  emissivity  computed using electric current and magnetic field, (although not obtained  in particular spetral lines as it would be with a realistic model) allows to characterise hot spots to compare with observations   (Figure~\ref{fig_simulation_emissivity}) .
%
While optical decoupling first occurs at the photosphere, beyond which photons freely escape, magnetic decoupling occurs above the Transition Region,  with TFRs being  a natural consequence. Ongoing and future investigations, especially with the Daniel K. Inouye Solar Telescope (DKIST) and Solar Orbiter}, are anticipated to deepen our understanding of those TFRs in the context of  atmospheric heating.

 %limitation
Finally, we would like to point out that although one limitation of the reconstruction made using  XTRAPOL in Section~\label{sec:reconstruction} is that it assumes a force-free hypothesis, which does not include the chromosphere and transition region as in the more realistic coupled MHD  simulation model discussed in this section, the  fact that these TFRs are also obtained in the coupled MHD simulation provides further evidence for the presence of TFRs. Furthermore, as the TFRs found lower down in the atmosphere might be more confined, the values of twist and energy relative to those of the semi open fields (that provide lower limits for eruption potential) for these TFRs might be even greater  than those usually found in pure force-free models. A more in-depth analysis of this simulation, including multiple other aspects beyond the TFRs presented here, will be presented in a forthcoming paper.

\begin{acknowledgments}
This work was granted access to the HPC resources of CINES/IDRIS under the allocation 2021-A0100500438 made by GENCI. The numerical simulations described in this paper have been performed on the HPE SGI 8600, Jean-Zay  of the institute I.D.R.I.S of the Centre National de la Recherche Scientifique,   through GENCI. We acknowledge support of the Centre National d\'Etudes Spatiales (CNES). 
% Computation:
% Temporary sentence and to be determined sooner
Ambiguity resolution and 3D diagnostics computations have been performed under the project n3s on the HPC facility Cholesky operated by the Ecole Polytechnique / IDCS.
% Data:
Hinode is a Japanese mission developed and launched by ISAS/JAXA, with NAOJ as domestic partner and NASA and STFC (UK) as international partners. It is operated by these agencies in cooperation with ESA and NSC (Norway).
\end{acknowledgments}

%\bibliography{mhd}{}

\begin{thebibliography}{}
\expandafter\ifx\csname natexlab\endcsname\relax\def\natexlab#1{#1}\fi
\providecommand{\url}[1]{\href{#1}{#1}}
\providecommand{\dodoi}[1]{doi:~\href{http://doi.org/#1}{\nolinkurl{#1}}}
\providecommand{\doeprint}[1]{\href{http://ascl.net/#1}{\nolinkurl{http://ascl.net/#1}}}
\providecommand{\doarXiv}[1]{\href{https://arxiv.org/abs/#1}{\nolinkurl{https://arxiv.org/abs/#1}}}

\bibitem[{{Aly} \& {Amari}(1985)}]{AlyAmari85}
{Aly}, J.~J., \& {Amari}, T. 1985, in Theoretical Problems in High Resolution
  Solar Physics, Vol. 212, MPA, Munchen, 319

\bibitem[{{Amari} \& {Aly}(2009)}]{AmariAl09}
{Amari}, T., \& {Aly}, J. 2009, in IAU Symposium, Vol. 257, IAU Symposium, ed.
  {N.~Gopalswamy \& D.~F.~Webb}, 211--222, \dodoi{10.1017/S1743921309029329}

\bibitem[{{Amari} \& {Aly}(1989)}]{AmariAl89}
{Amari}, T., \& {Aly}, J.~J. 1989, \aap, 208, 261

\bibitem[{{Amari} \& {Aly}(2010)}]{AmariAl10}
---. 2010, \aap, 522, A52, \dodoi{10.1051/0004-6361/200913058}

\bibitem[{{Amari} {et~al.}(2014){Amari}, {Canou}, \& {Aly}}]{AmariCaAl2014}
{Amari}, T., {Canou}, A., \& {Aly}, J.-J. 2014, \nat, 514, 465,
  \dodoi{10.1038/nature13815}

\bibitem[{{Amari} {et~al.}(2018){Amari}, {Canou}, {Aly}, {Delyon}, \&
  {Alauzet}}]{AmariAlCa2018}
{Amari}, T., {Canou}, A., {Aly}, J.-J., {Delyon}, F., \& {Alauzet}, F. 2018,
  \nat, 554, 211, \dodoi{10.1038/nature24671}

\bibitem[{{Amari} {et~al.}(2004){Amari}, {Luciani}, \& {Aly}}]{AmariLuAl04}
{Amari}, T., {Luciani}, J.~F., \& {Aly}, J.~J. 2004, \apjl, 615, L165

\bibitem[{{Amari} {et~al.}(2005){Amari}, {Luciani}, \& {Aly}}]{AmariLuAl05}
---. 2005, \apjl, 629, L37

\bibitem[{{Amari} {et~al.}(2008){Amari}, {Luciani}, \& {Aly}}]{AmariLuAl08}
---. 2008, \apjl, 681, L45, \dodoi{10.1086/590323}

\bibitem[{{Amari} {et~al.}(2015){Amari}, {Luciani}, \& {Aly}}]{AmariLuAl2015}
{Amari}, T., {Luciani}, J.-F., \& {Aly}, J.-J. 2015, \nat, 522, 188,
  \dodoi{10.1038/nature14478}

\bibitem[{{Anzer} \& {Priest}(1985)}]{AnzerPri85}
{Anzer}, U., \& {Priest}, E. 1985, \solphys, 95, 263

\bibitem[{{Archontis} \& {T{\"o}r{\"o}k}(2008)}]{ArchontisTor2008}
{Archontis}, V., \& {T{\"o}r{\"o}k}, T. 2008, \aap, 492, L35,
  \dodoi{10.1051/0004-6361:200811131}

\bibitem[{{Aschwanden} {et~al.}(2007){Aschwanden}, {Winebarger}, {Tsiklauri},
  \& {Peter}}]{Aschwandenetal2007}
{Aschwanden}, M.~J., {Winebarger}, A., {Tsiklauri}, D., \& {Peter}, H. 2007,
  \apj, 659, 1673, \dodoi{10.1086/513070}

\bibitem[{{Aulanier}(2014)}]{Aulanier2014}
{Aulanier}, G. 2014, in Nature of Prominences and their Role in Space Weather,
  ed. B.~{Schmieder}, J.-M. {Malherbe}, \& S.~T. {Wu}, Vol. 300, 184--196,
  \dodoi{10.1017/S1743921313010958}

\bibitem[{{Aulanier} \& {D\'emoulin}(1998)}]{AulanierDe98}
{Aulanier}, G., \& {D\'emoulin}, P. 1998, \aap, 329, 1125

\bibitem[{{Berghmans} {et~al.}(2021){Berghmans}, {Auch{\`e}re}, {Long}, \&
  et~al.}]{BerghamsAuetal2021}
{Berghmans}, D., {Auch{\`e}re}, F., {Long}, D.~M., \& et~al. 2021, Astronomy \&
  Astrophysics, 656, A79, \dodoi{10.1051/0004-6361/202140380}

\bibitem[{Bose {et~al.}(2024)Bose, De~Pontieu, Hansteen, {et~al.}}]{Bose2024}
Bose, S., De~Pontieu, B., Hansteen, V., {et~al.} 2024, Nature Astronomy,
  \dodoi{10.1038/s41550-024-02241-8}

\bibitem[Bushby \& Favier(2014)]{Bushby2014} Bushby, P.~J. \& Favier, B.\ 2014, \aap, 562, A72. doi:10.1051/0004-6361/201322993

\bibitem[{{Chen} {et~al.}(2021){Chen}, {Przybylski}, {Peter}, {Tian},
  {Auchere}, \& {Berghmans}}]{ChenPeter2021}
{Chen}, Y., {Przybylski}, D., {Peter}, H., {et~al.} 2021, \aap, 656, L7,
  \dodoi{10.1051/0004-6361/202140638}

\bibitem[{{Cheung} \& {DeRosa}(2012)}]{Cheungetal2012}
{Cheung}, M. C.~M., \& {DeRosa}, M.~L. 2012, \apj, 757, 147,
  \dodoi{10.1088/0004-637X/757/2/147}

\bibitem[{{Cheung} {et~al.}(2019){Cheung}, {Rempel}, {Chintzoglou}, {Chen},
  {Testa}, {Mart{\'\i}nez-Sykora}, {Sainz Dalda}, {DeRosa}, {Malanushenko},
  {Hansteen}, {De Pontieu}, {Carlsson}, {Gudiksen}, \&
  {McIntosh}}]{Cheungetal2019}
{Cheung}, M.~C.~M., {Rempel}, M., {Chintzoglou}, G., {et~al.} 2019, Nature
  Astronomy, 3, 160, \dodoi{10.1038/s41550-018-0629-3}

\bibitem[Del Zanna et al.(2021)]{CHIANTI10} Del Zanna, G., Dere, K.~P., Young, P.~R., et al.\ 2021, \apj, 909, 38. doi:10.3847/1538-4357/abd8ce

\bibitem[{{D\'emoulin} \& {Priest}(1988)}]{DemoulinPriest88}
{D\'emoulin}, P., \& {Priest}, E.~R. 1988, \aap, 206, 336

\bibitem[Demoulin et al.(1996)]{Demoulin1996} Demoulin, P., Henoux, J.~C., Priest, E.~R., et al.\ 1996, \aap, 308, 643

\bibitem[Dere et al.(1997)]{CHIANTI1} Dere, K.~P., Landi, E., Mason, H.~E., et al.\ 1997, \aaps, 125, 149. doi:10.1051/aas:1997368

\bibitem[{{Fan}(2022)}]{Fan2022}
{Fan}, Y. 2022, \apj, 941, 61, \dodoi{10.3847/1538-4357/aca0ec}

\bibitem[{{Fan} \& {Gibson}(2007)}]{FanGib07}
{Fan}, Y., \& {Gibson}, S. 2007, \apj, 668, 1232, \dodoi{10.1086/521335}

\bibitem[{{Fan} \& {Gibson}(2004)}]{FanGibson04}
{Fan}, Y., \& {Gibson}, S.~E. 2004, \apj, 609, 1123

\bibitem[{{Gary} \& {Hagyard}(1990)}]{Garyetal1990}
{Gary}, G.~A., \& {Hagyard}, M.~J. 1990, \solphys, 126, 21,
  \dodoi{10.1007/BF00158295}

\bibitem[{{Go{\v{s}}i{\'c}} {et~al.}(2024){Go{\v{s}}i{\'c}}, {De Pontieu}, \&
  {Sainz Dalda}}]{Gosicetal2024}
{Go{\v{s}}i{\'c}}, M., {De Pontieu}, B., \& {Sainz Dalda}, A. 2024, \apj, 964,
  175, \dodoi{10.3847/1538-4357/ad2e03}

\bibitem[{{Kliem} \& {T{\"o}r{\"o}k}(2006)}]{KliemTo06}
{Kliem}, B., \& {T{\"o}r{\"o}k}, T. 2006, Physical Review Letters, 96, 255002,
  \dodoi{10.1103/PhysRevLett.96.255002}

\bibitem[{Kosugi {et~al.}(2007)Kosugi, Matsuzaki, Sakao, Shimizu, Sone,
  Tachikawa, Hashimoto, Minesugi, Ohnishi, Yamada, Tsuneta, Hara, Ichimoto,
  Suematsu, Shimojo, Watanabe, Shimada, Davis, Hill, Owens, Title, Culhane,
  Harra, Doschek, \& Golub}]{Kosugi2007}
Kosugi, T., Matsuzaki, K., Sakao, T., {et~al.} 2007, Solar Physics, 243, 3,
  \dodoi{10.1007/s11207-007-9014-6}

\bibitem[{{Kuperus} \& {Raadu}(1974)}]{KuperusRaadu1974}
{Kuperus}, M., \& {Raadu}, M.~A. 1974, \aap, 31, 189

\bibitem[{{Leake} {et~al.}(2022){Leake}, {Linton}, \&
  {Antiochos}}]{LeakeLinAn2022}
{Leake}, J.~E., {Linton}, M.~G., \& {Antiochos}, S.~K. 2022, \apj, 934, 10,
  \dodoi{10.3847/1538-4357/ac74b7}

\bibitem[{{Leka} {et~al.}(2009){Leka}, {Barnes}, {Crouch}, {Metcalf}, {Gary},
  {Jing}, \& {Liu}}]{Lekaetal2009}
{Leka}, K.~D., {Barnes}, G., {Crouch}, A.~D., {et~al.} 2009, \solphys, 260, 83,
  \dodoi{10.1007/s11207-009-9440-8}

\bibitem[{{Lites} {et~al.}(2008){Lites}, {Kubo}, {Socas-Navarro}, {Berger},
  {Frank}, {Shine}, {Tarbell}, {Title}, {Ichimoto}, {Katsukawa}, {Tsuneta},
  {Suematsu}, {Shimizu}, \& {Nagata}}]{Lites2008}
{Lites}, B.~W., {Kubo}, M., {Socas-Navarro}, H., {et~al.} 2008, \apj, 672,
  1237, \dodoi{10.1086/522922}

\bibitem[{{L{\'o}pez Fuentes} {et~al.}(2000){L{\'o}pez Fuentes}, {D\'emoulin},
  {Mandrini}, \& {van Driel-Gesztelyi}}]{LopenzFuentesDeManVan00}
{L{\'o}pez Fuentes}, M.~C., {D\'emoulin}, P., {Mandrini}, C.~H., \& {van
  Driel-Gesztelyi}, L. 2000, \apj, 544, 540, \dodoi{10.1086/317180}

\bibitem[{{Low}(2001)}]{Low01}
{Low}, B.~C. 2001, \jgr, 106, 25141, \dodoi{10.1029/2000JA004015}

\bibitem[{{MacTaggart} {et~al.}(2021){MacTaggart}, {Prior}, {Raphaldini},
  {Romano}, \& {Guglielmino}}]{MacTaggartetal2021}
{MacTaggart}, D., {Prior}, C., {Raphaldini}, B., {Romano}, P., \&
  {Guglielmino}, S.~L. 2021, Nature Communications, 12, 6621,
  \dodoi{10.1038/s41467-021-26981-7}

\bibitem[{{Mart{\'\i}nez-Sykora} {et~al.}(2019){Mart{\'\i}nez-Sykora}, {De
  Pontieu}, {Hansteen}, {Rouppe van der Voort}, {Carlsson}, \&
  {Pereira}}]{MartinezSikora2019}
{Mart{\'\i}nez-Sykora}, J., {De Pontieu}, B., {Hansteen}, V.~H., {et~al.} 2019,
  \apj, 878, 40, \dodoi{10.3847/1538-4357/ab1f0b}

\bibitem[{{Panesar} {et~al.}(2021){Panesar}, {Tiwari}, {Berghmans}, {Cheung},
  {M{\"u}ller}, {Auchere}, \& {Zhukov}}]{Panesaretal2021}
{Panesar}, N.~K., {Tiwari}, S.~K., {Berghmans}, D., {et~al.} 2021, \apjl, 921,
  L20, \dodoi{10.3847/2041-8213/ac3007}

\bibitem[Parker(1983)]{Parker1983} Parker, E.~N.\ 1983, \apj, 264, 635. doi:10.1086/160636

\bibitem[{Patsourakos {et~al.}(2020)Patsourakos, Vourlidas, T{\"o}r{\"o}k,
  Kliem, \& Antiochos}]{Patsourakos2020}
Patsourakos, S., Vourlidas, A., T{\"o}r{\"o}k, T., Kliem, B., \& Antiochos,
  S.~K. 2020, Space Science Reviews, 216

\bibitem[{{Priest} \& {Forbes }(2002)}]{PriestForbes02}
{Priest}, E.~R., \& {Forbes }, T.~G. 2002, Astron Astrophys Rev, 10, 313

\bibitem[{{R{\'e}gnier} {et~al.}(2008){R{\'e}gnier}, {Parnell}, \&
  {Haynes}}]{Regnier2008}
{R{\'e}gnier}, S., {Parnell}, C.~E., \& {Haynes}, A.~L. 2008, \aap, 484, L47,
  \dodoi{10.1051/0004-6361:200809826}

\bibitem[{Robinson {et~al.}(2023)Robinson, Aulanier, \&
  Carlsson}]{Robinson2023}
Robinson, R., Aulanier, G., \& Carlsson, M. 2023, Astronomy \& Astrophysics,
  673, A79, \dodoi{10.1051/0004-6361/202346065}

\bibitem[{{Schmieder}(2022)}]{Schmieder2022}
{Schmieder}, B. 2022, Frontiers in Astronomy and Space Sciences, 9, 820183,
  \dodoi{10.3389/fspas.2022.820183}

\bibitem[{{Schrijver} \& {Title}(2003)}]{Schrijver2003}
{Schrijver}, C.~J., \& {Title}, A.~M. 2003, \apjl, 597, L165,
  \dodoi{10.1086/379870}

\bibitem[{{Skumanich} \& {Lites}(1987)}]{Skumanichetal1987}
{Skumanich}, A., \& {Lites}, B.~W. 1987, \apj, 322, 473, \dodoi{10.1086/165743}

\bibitem[{{Titov}(2007)}]{Titov2007}
{Titov}, V.~S. 2007, \apj, 660, 863, \dodoi{10.1086/512671}

\bibitem[{{T{\"o}r{\"o}k} \& {Kliem}(2005)}]{TorokKl05}
{T{\"o}r{\"o}k}, T., \& {Kliem}, B. 2005, \apjl, 630, L97,
  \dodoi{10.1086/462412}

\bibitem[{{T{\"o}r{\"o}k} {et~al.}(2018){T{\"o}r{\"o}k}, {Downs}, {Linker},
  {Lionello}, {Titov}, {Miki{\'c}}, {Riley}, {Caplan}, \&
  {Wijaya}}]{Toroketal2018}
{T{\"o}r{\"o}k}, T., {Downs}, C., {Linker}, J.~A., {et~al.} 2018, \apj, 856,
  75, \dodoi{10.3847/1538-4357/aab36d}

\bibitem[{{Trujillo Bueno} {et~al.}(2004){Trujillo Bueno}, {Shchukina}, \&
  {Asensio Ramos}}]{TrujilloBueno2004}
{Trujillo Bueno}, J., {Shchukina}, N., \& {Asensio Ramos}, A. 2004, \nat, 430,
  326, \dodoi{10.1038/nature02669}

\bibitem[{Tsuneta {et~al.}(2008)Tsuneta, Ichimoto, Katsukawa, Nagata, Otsubo,
  Shimizu, Suematsu, Nakagiri, Noguchi, Tarbell, Title, Shine, Rosenberg,
  Hoffmann, Jurcevich, Kushner, Levay, Lites, Elmore, Matsushita, Kawaguchi,
  Saito, Mikami, Hill, \& Owens}]{Tsuneta2008}
Tsuneta, S., Ichimoto, K., Katsukawa, Y., {et~al.} 2008, Solar Physics, 249,
  167, \dodoi{10.1007/s11207-008-9174-z}

\bibitem[{{van Ballegooijen} \& {Martens}(1989)}]{VanBallegooijenMa89}
{van Ballegooijen}, A.~A., \& {Martens}, P.~C.~H. 1989, \apj, 343, 971,
  \dodoi{10.1086/167766}

\bibitem[{{Zhukov} {et~al.}(2021){Zhukov}, {Mierla, M.}, {Auchere, F.},
  {Gissot, S.}, {Rodriguez, L.}, {Soubri, E.}, {Thompson, W. T.}, {Inhester,
  B.}, {Nicula, B.}, {Antolin, P.}, {Parenti, S.}, {Buchlin, Ã.}, {Barczynski,
  K.}, {Verbeeck, C.}, {Kraaikamp, E.}, {Smith, P. J.}, {Stegen, K.}, {Dolla,
  L.}, {Harra, L.}, {Long, D. M.}, {SchÃŒhle, U.}, {Podladchikova, O.}, {Aznar
  Cuadrado, R.}, {Teriaca, L.}, {Haberreiter, M.}, {Katsiyannis, A. C.},
  {Rochus, P.}, {Halain, J.-P.}, {Jacques, L.}, \& {Berghmans,
  D.}}]{Zukhovetal2021}
{Zhukov}, {Mierla, M.}, {Auchere, F.}, {et~al.} 2021, \aap, 656, A35,
  \dodoi{10.1051/0004-6361/202141010}

\end{thebibliography}
%\bibliographystyle{aasjournal}

%

%\end{document}
%\end{document}

\clearpage

\clearpage

\clearpage

\begin{figure}
\plotone{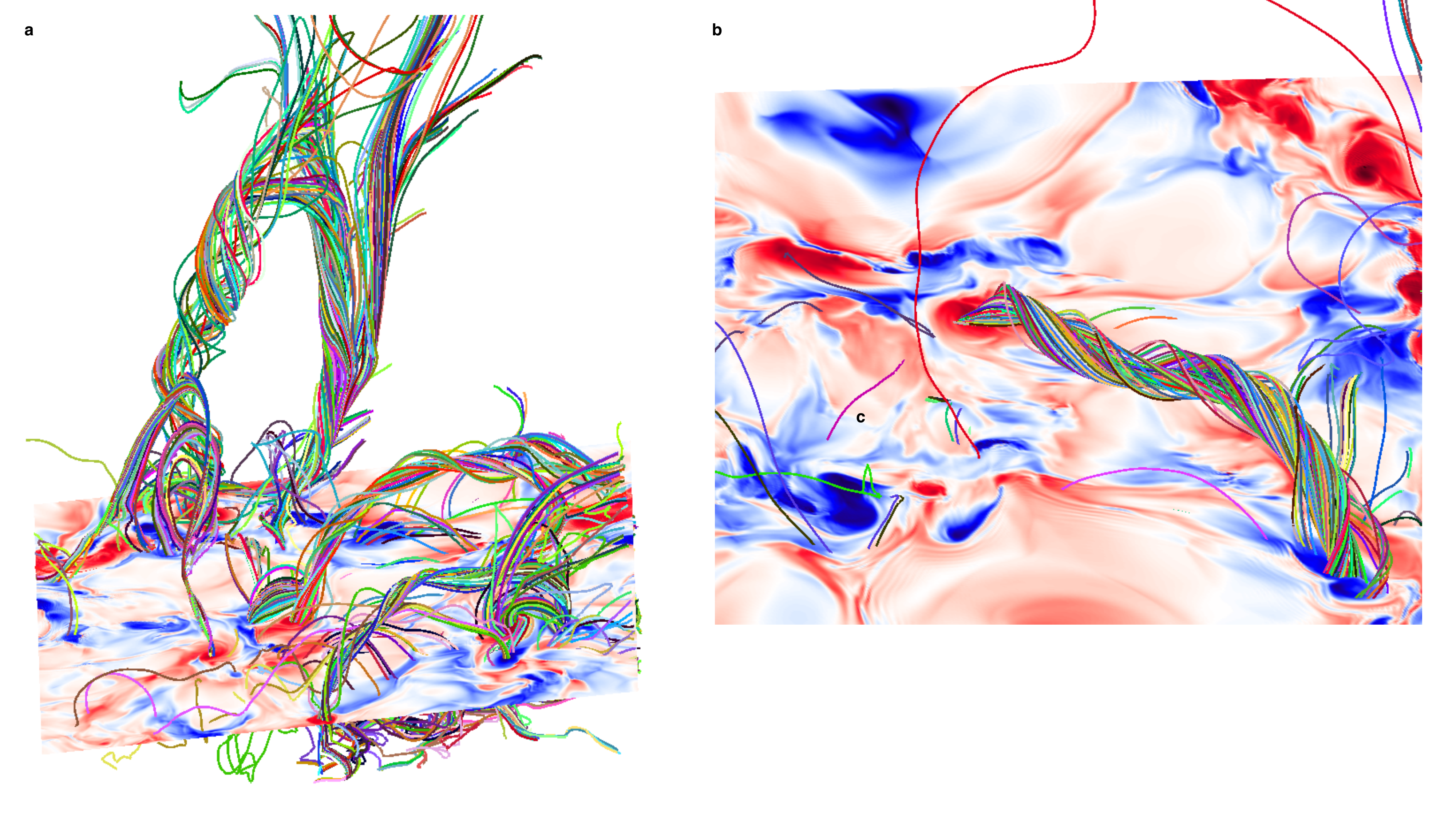}
\epsscale{1.}
%\plotone{fig1.eps}
\caption{Selected field lines of the magnetic environment above the photosphere during the simulation, taken at two different times and two different view angles. Twisted Flux Ropes (TFRs) around mesospots and below and above the transition regions. The horizontal size of the domain is the same as that of the smallest rectangle drawn in green in Figure~\ref{fig_reconst_bzjz}. For both views a  horizontal scalar map of the vertical component of the magnetic field in the high chromosphere.
\label{fig_simulation_tfr}}
\end{figure}
\clearpage

\end{document}